\documentclass{jsara}
\journalinfo{Journal of the Southeastern Association for Research in Astronomy, {\bf VOL}, 5, 2012}

\slugcomment{}

\shorttitle{Variable Stars in M14}
\shortauthors{Conroy et al.}

\begin{document}

\setcounter{page}{34}


\title{Variable Stars in the Globular Cluster M14}


\author{Kyle E. Conroy\altaffilmark{1}}
\affil{Department of Astronomy and Astrophysics, Villanova University, 
Villanova, PA, 19085}
\and
\author{Andrew N. Darragh, Zheyu J. Liu, and Brian W. Murphy}
\affil{Department of Physics and Astronomy, Butler University, 
Indianapolis, IN, 46208}

\altaffiltext{1}{Southeastern Association for Research in Astronomy
(SARA) NSF-REU Summer Intern}
\email{bmurphy@butler.edu (BWM)}

\begin{abstract}
Using an image subtraction method  we have searched for variable stars
in  the  globular  cluster  M14.   We confirmed  62  previously  known
variables catalogued by Wehlau \& Froelich (1994).  In addition to the
previously known  variables we have  identified 71 new  variables.  We
have confirmed the  periods of most of Wehlau  \& Froelich's variables
we found with just a few exceptions.  Of the total number of confirmed
variables, we  found a total of  112 RR Lyrae stars,  several of which
exhibited the Blazhko  Effect.  Of the total we  classified 55 RR0, 57
RR1, 19  variables with periods greater  than 2 days, a  W UMa contact
binary,  and an SX  Phe star.   We present  the periods  of previously
found  variables as  well as  the periods,  classification,  and light
curves of the newly discovered variables.
\end{abstract}


\keywords{stars:   variables:   general--Galaxy:  globular   clusters:
  individual: M14}


\section{Introduction}

Globular clusters provide a  laboratory for studying stellar evolution
of a coeval  population of stars.  This allows  for the examination of
various evolutionary phases of  the stellar population.  Of particular
interest are  post main sequence variable stars  in globular clusters.
The most  common type  of variable stars  in globular clusters  are RR
Lyrae stars.  These stars are  horizontal branch stars that lie within
the instability strip.  Only low metallicity clusters contain RR Lyrae
stars.  Having a metallicity  of [Fe/H]=-1.39 the globular cluster M14
is one such cluster (Harris 1996).

M14 (NGC 6402)  is a globular cluster with  many known variable stars.
Clement's   Catalogue   of  Variable   Stars   in  Globular   Clusters
\citep{cle2001}  reports  90  possible  variables  of  which  61  have
determined periods.  54 of these  have been identified as RR Lyraes, 6
as Cepheids or  RV Tau, 6 as long term or  irregular variables, and no
eclipsing binaries or  SX Phoenix stars.  Most of  these were found by
\citet{weh1994} using photographic photometry on data obtained between
1912 and 1980 mostly  by H.  S. Hogg \citep{sawyerhogg68}.  Additional
unpublished studies  have also been  conducted which do not  appear in
Clement's  catalogue,  and  have  found several  additional  variables
(Jacobs 2004).  M14  lies near the celestial equator  so is visible in
both the  southern and northern hemisphere. Given  our extended access
to telescopes in each hemisphere, M14 was an ideal candidate to search
for variables.

In  this  study  we  use  an image  subtraction  method  developed  by
\citet{ala2000}  to   search  the  central   $13\times13{\arcmin}$  of
globular cluster M14 for  variable stars from observations obtained in
June and July of 2010.   With the combination of better resolution CCD
images and  image subtraction, we can better  resolve variables dimmer
and/or closer to the crowded field in the core as compared to previous
photographic  studies.  In  doing so,  we have  confirmed many  of the
previously  found  variables,  confirmed  or  have  better  determined
periods, as well as identified a  large number of new variables in the
cluster.

\section{Observations and Reduction}

Image frames were obtained using two different telescopes, one at Kitt
Peak  National  Observatory  (KPNO)  and  the other  at  Cerro  Tololo
Interamerican Observatory (CTIO).   On the nights of 6,  9, 16, 18 and
19 June 2010,  images were obtained using the  KPNO SARA (Southeastern
Association  for Research in  Astronomy) 0.9  meter telescope  with an
Apogee Alta U42  CCD with a 2048$\times$2048 Kodak  e2V CC42-40 with a
gain  of 1.2  electrons per  count, RMS  noise of  6.3  electrons, and
cooled  to  a  temperature   of  approximately  -30  degrees  Celsius.
1$\times$1 binning was used,  resulting in a scale of $0.42\arcsec/$px
and  a  $13.6\times13.6\arcmin$ field  of  view.   Typical seeing  was
2.2$\arcsec$  and  ranged from  1.5-3.0$\arcsec$.   Using  a Bessel  R
filter,  exposure times were  set at  a constant  60 seconds  to avoid
overexposure of the core or  bright giants during best possible seeing
conditions.

Another set of images were taken  with the SARA 0.6 meter telescope at
CTIO on 3, 4, 5, 9, 13, and  19 July 2010 using an Apogee Alta E6 with
a 1024$\times$1024  Kodak KAF1001E chip,  with a gain of  1.5 electron
per count, and an RMS noise of 8.9 electrons. The temperature was held
at -30  Celsius.  1$\times$1 binning  was used with a  resulting image
scale  of $0.6\arcsec/$px  and  a $10\times10\arcmin$  field of  view.
Typical seeing  was 1.8$\arcsec$ and ranged from  1.2 to 2.5$\arcsec$.
Using a  Bessel V filter,  exposure times were  set at a  constant 150
seconds to achieve similar images as with the KPNO observations.  This
also ensured  that none  of the bright  stars were  overexposed during
periods of good seeing.  Maxim DL  was used to process the CCD images.
Each  image was  debiased, flat-fielded,  and  dark-subtracted.  These
processed images were then analyzed using image subtraction.

The  autoguider at  KPNO was  not functioning  properly except  on the
night of 6 June, and was also not functioning properly for a number of
nights  but the  CTIO  telescope  tracked much  better  than the  KPNO
telescope minimizing the importance for autoguiding.  Because of this,
the CTIO telescope resulted in higher quality images.  The KPNO images
were   only  used   to  examine   regions  outside   of   the  smaller
$10\times10\arcmin$ field of view of the CTIO telescope.

\section{Image Subtraction}

Image   Subtraction   was  completed   using   the  ISIS-2.1   package
\citep{ala1998,ala2000}.  The  ISIS package consists  of six different
c-shell scripts:  {\tt interp} (registration  and interpolation), {\tt
  ref} (builds reference frame), {\tt subtract} (subtracts images from
reference), {\tt detect} (stacks subtracted images), {\tt find} (finds
variables above  user-defined threshold), and {\tt  phot} (makes light
curves for found variables), along with three parameter files.

ISIS accounts  for changes in  seeing conditions by convolving  a high
quality  reference  image  to  the  point  spread  functions  of  each
individual frame and  then subtracting the two images  to determine if
any  change in intensity  has occurred.   These subtracted  frames are
then combined to create a  {\tt var.fits} image which shows the degree
of  variability  across the  observation  run  for  each star  in  the
cluster.

The {\tt SIGTHRESH} parameter controls  the threshold for which a star
is considered a  variable. Typical values were for  0.15 for SARA KPNO
and 0.05 for  SARA CTIO observations, with each  producing roughly 270
possible  variables.   Many of  which  were  due  to noise  and  later
eliminated after  examining the  individual light curves.   Values for
{\tt  SIGTHRESH} were  found by  looking for  the point  at  which the
number of possible objects increased rapidly.  A lower {\tt SIGTHRESH}
value typically led  to a thousand or more  false-positives.  Once the
proper {\tt SIGTHRESH} value was found and false positives eliminated,
light curves  of relative flux  were then  generated for  each detected
variable.

In  order to maintain  consistent and  comparable values  for relative
flux across nights, one reference  image was created using five of the
best images  from June 6 and this  was then used as  the reference for
all of the KPNO observations.   Since the CCD pixel scale is different
on the  CTIO telescope, a different  reference image was  used for the
CTIO observations.   Thirteen of the images with  best seeing (average
of 1.4$\arcsec$) were used to  create the reference image used for all
CTIO observations.  The {\tt  ref.fits} and {\tt var.fits} images from
SARA Cerro Tololo are shown in Figures 1 and 2.

\begin{figure}
\epsscale{1.15}
\plotone{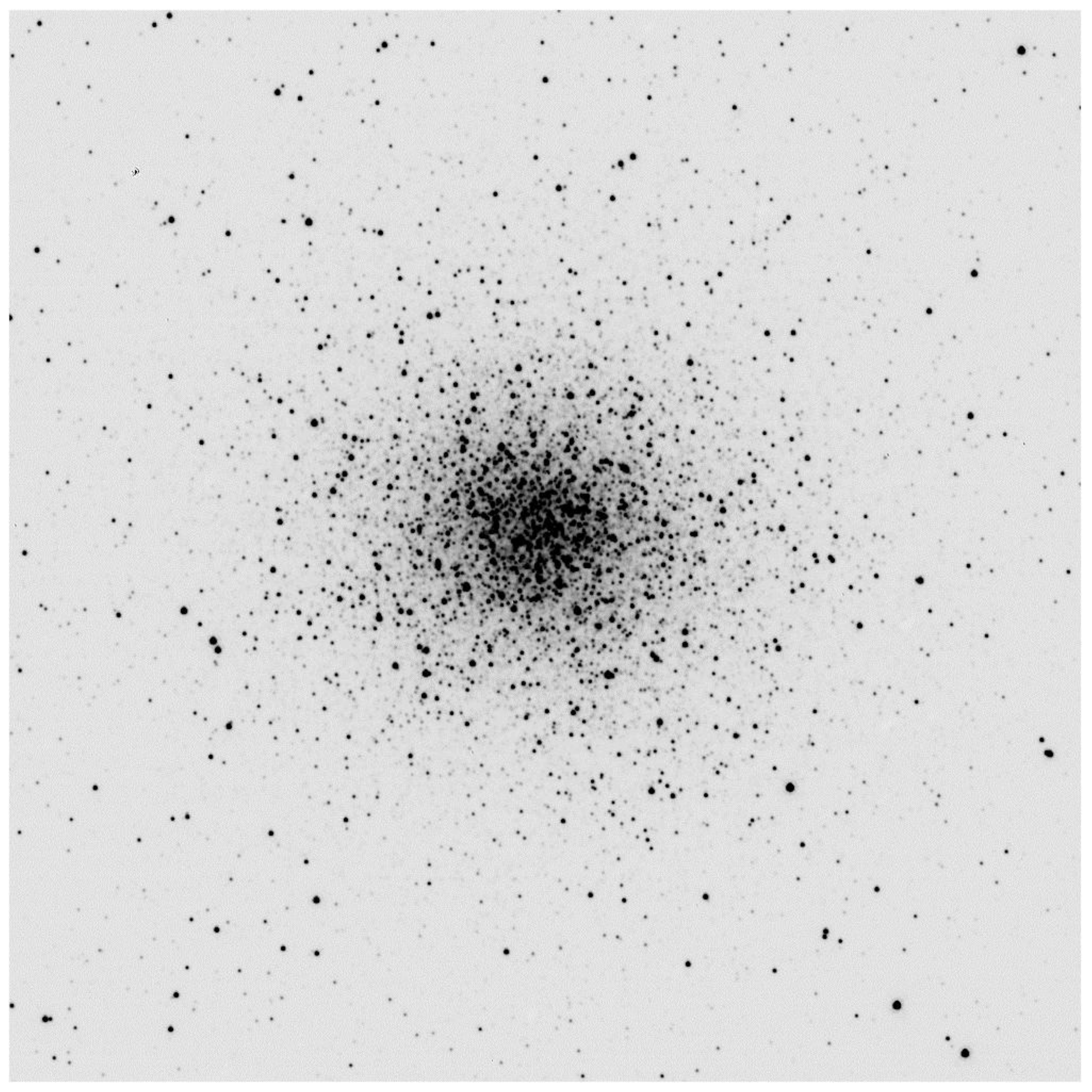}
\caption{The {\tt ref.fits} used for  all nights from SARA CTIO.  This
  image is a combination of  the best seeing images, with the combined
  seeing near $1.4\arcsec$. }
\label{reffits}
\end{figure}

\begin{figure}
\epsscale{1.15}
\plotone{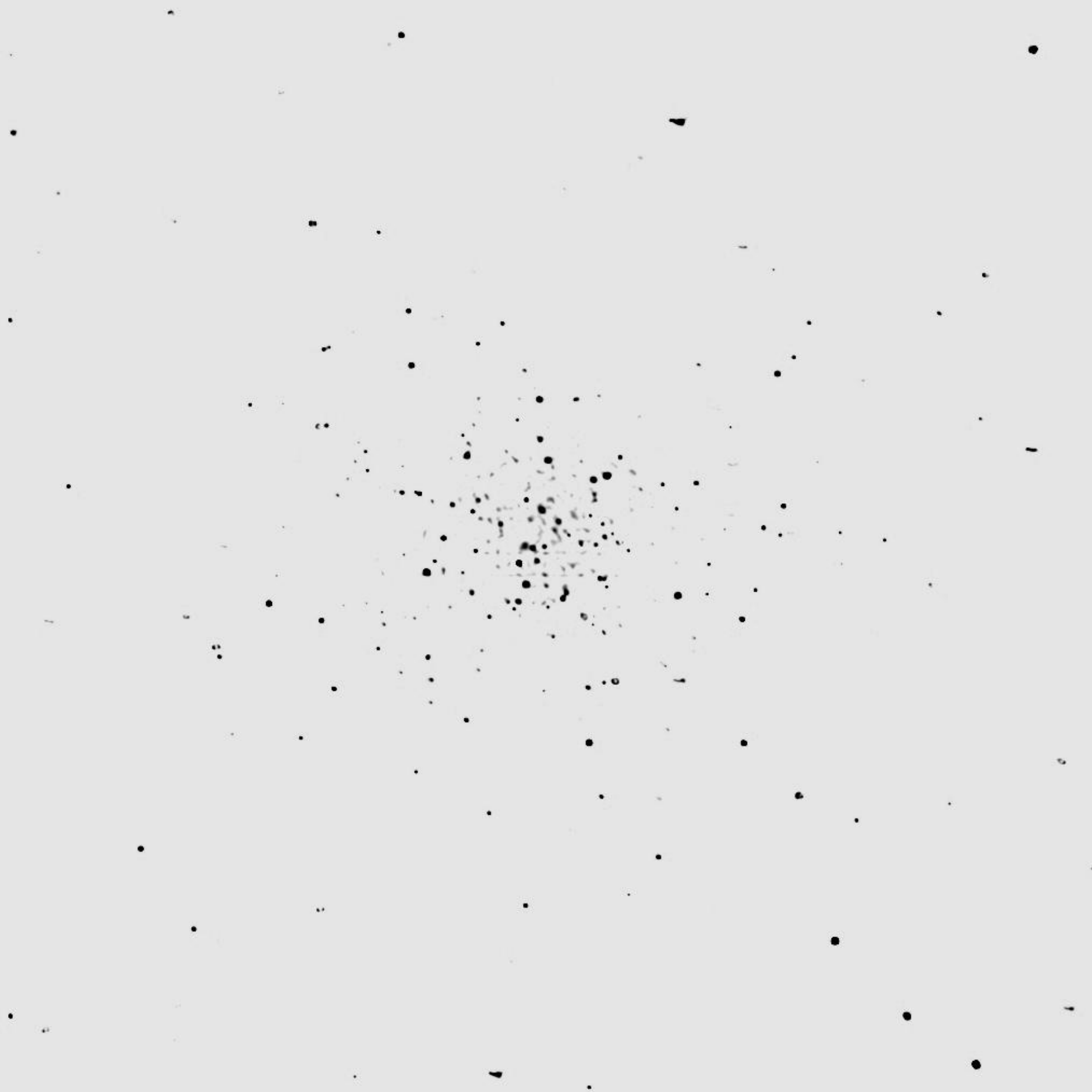}
\caption{The {\tt var.fits} from ISIS for all six nights from SARA
  CTIO.  The relative amount of variation is indicated by the
  brightness of the star.}
\label{varfits}
\end{figure}

\section{Analysis}
\subsection{Periods}
Periods were determined, when  possible, based on combined photometric
data from the runs on each  of the two telescopes.  We used the period
finding software {\tt AVE}  (Analisis de Variabilidad Estela, Analysis
of  Estellar  Variability)   from  Grup  d'Estudis  Astronomics.   The
software uses the Lomb-Scargle algorithm \citep{scargle82}.  With data
over six full nights, we were  able to determine periods of nearly all
of the variables having periods under 3 days as can be seen in Table 1
\&  2.   However, a  few  of  these  variables had  mulitple  possible
periods.  The periods were  generally accurate to $10^{-4}$ days.  Due
to  our lack  of long  term obserrvations  spanning several  months or
more, we were unable to  accurately ascertain the periods of variables
greater than a  few days.  As we obtain more  observations, we hope to
determine the periods of these variables as well.

Apart from  the longer  period variables, we  verified the  periods of
nearly all of the Wehlau \&  Froelich (1994) variables with only a few
exceptions where  there were mulitple  periods possible.  The  only RR
Lyrae star that we found a significant period differing from Wehlau \&
Froelich was 77.  We found a period of 0.7910 versus 0.3274 days.

\begin{figure}
\epsscale{1.3}
\plotone{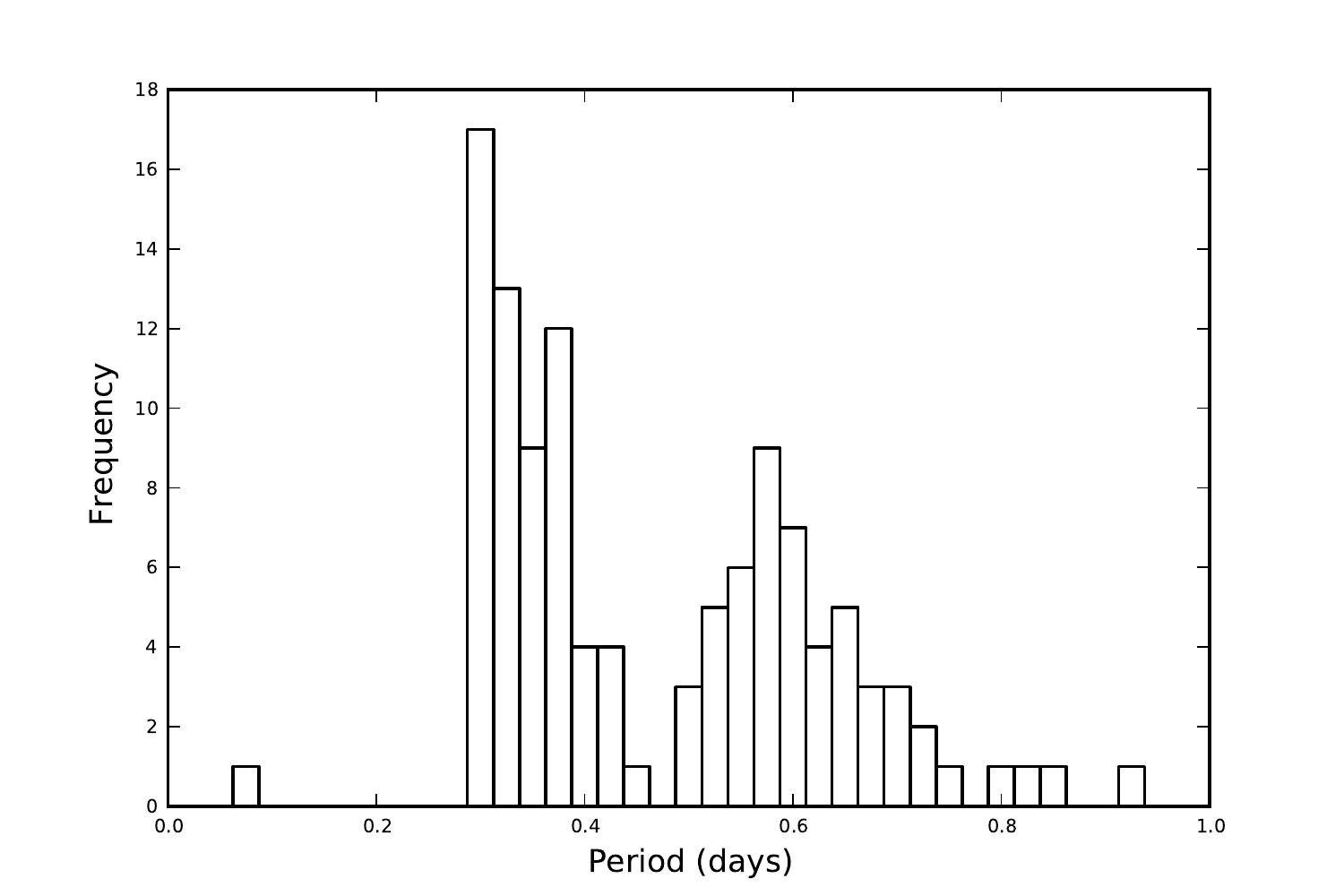}
\caption{A period  histogram of  observed variables with  periods less
  than 1 day.   Note the distinctive break just  above 0.4 days.  This
  break separates  the RR0  variable stars (right  peak) from  the RR1
  variables (left peak).}
\label{period_hist}
\end{figure}

\begin{figure}
\epsscale{1.3}
\plotone{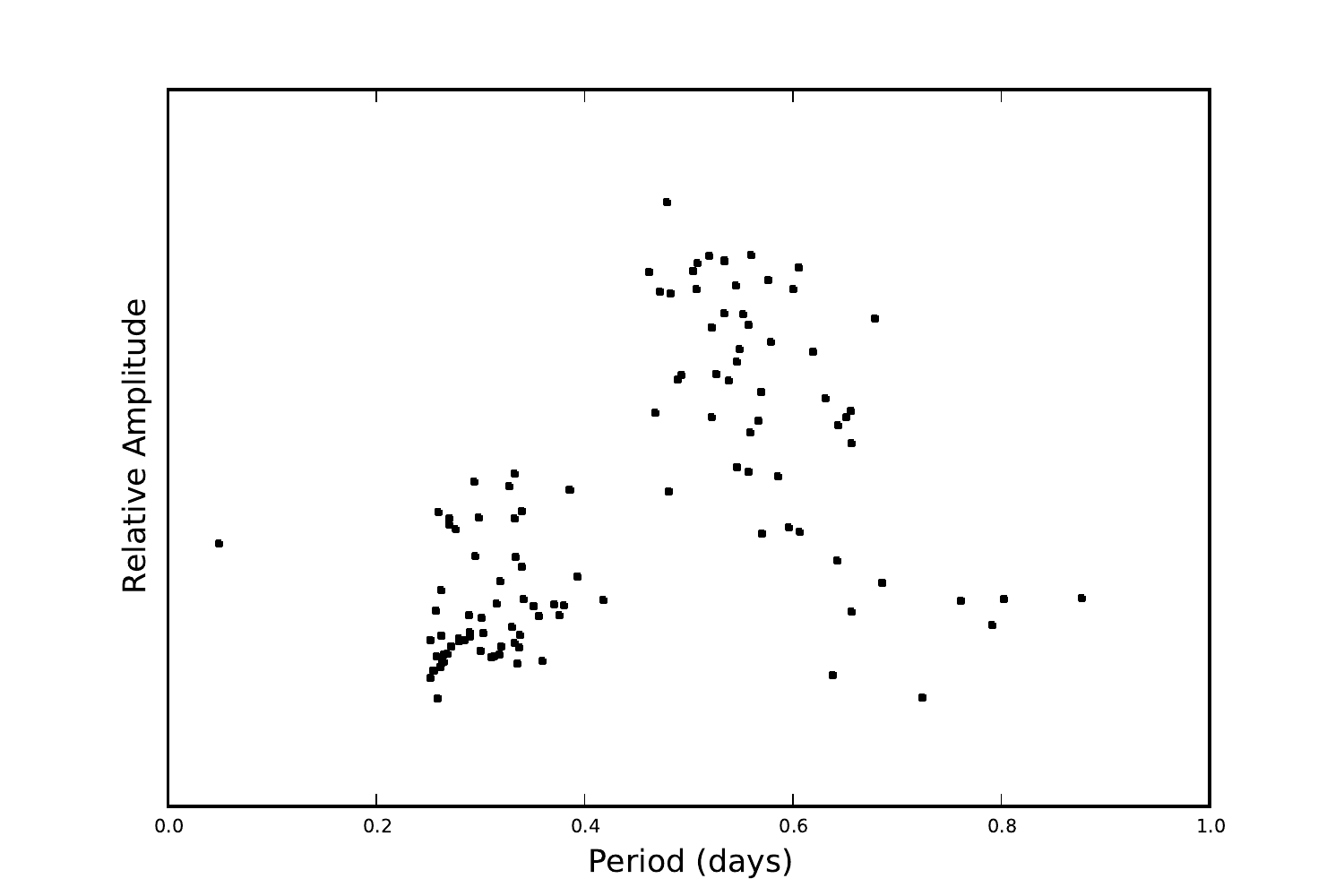}
\caption{A  plot  of variable  period  versus  amplitude.  Similar  to
  figure  3 there  is  a distinctive  separation  of the  RR0 and  RR1
  stars.}
\label{period_amp}
\end{figure}

\subsection{Identification}

Astrometry was done  by finding a plate solution  using positions from
Clement's   Catalogue   of  Variable   Stars   in  Globular   Clusters
\citep{cle2001}.  We then  found additional potential variable matches
by  comparing our  coordinates  with those  in previous  publications.
Once these were found, they were confirmed through period comparisons.
We compared our results with  Wehlau and to the unpublished results of
Jacobs (2004).   Unfortunately the Jacobs  astrometry had an  error in
the right ascension  so we could only make a  useful comparison to the
Wehlau \& Froelich variables in the Jacobs' data set.

In  keeping with  previously  published work,  we  retained Wehlau  \&
Froelich's  number system for  stars 1  through 93.   Newly discovered
variables  from  this  research  are  in  order  of  increasing  right
ascension and begin with variable  number 94.  Of Wehlau \& Froelich's
93 variables,  we were able to find  and confirm 63 of  them.  Many of
those  variables we could  not confirm  within our  field of  view are
listed  as  {\it  blended}  or  {\it NV}  (meaning  not  variable)  in
\citet{cle2001},  and  thus  their  variability is  either  suspect  or
difficult to  determine.  Two others  variables (V27 and 28)  were not
observed because  they lay  outside of our  field of view.   This left
only six  remaining variables  that we were  unable to confirm  in the
\citet{cle2001} catalogue (V25, 45, 68, 69, 71 and 86).

\begin{figure}
\epsscale{1.15}
\plotone{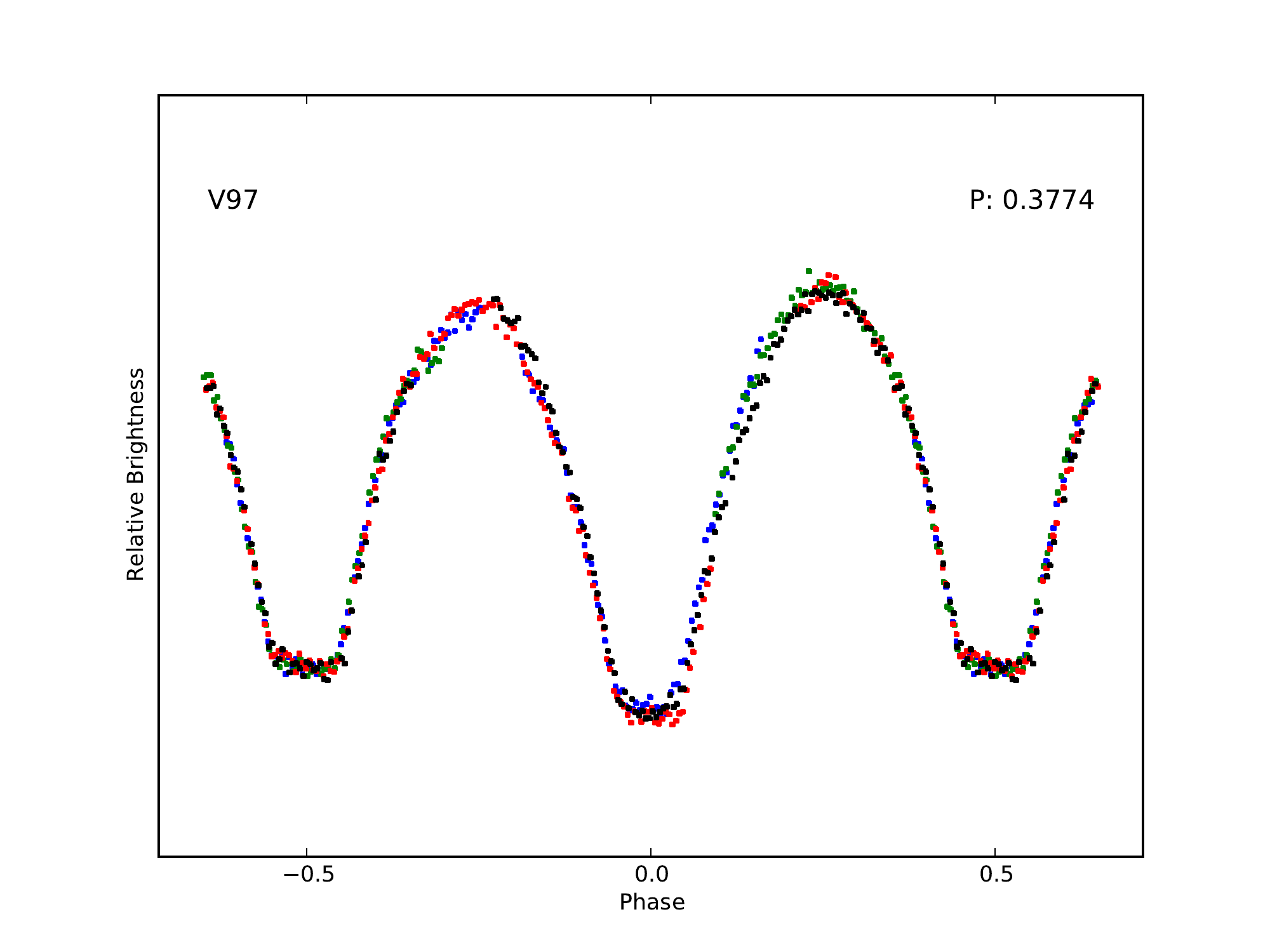}
\caption{Phased  light  curve  of  the  over-contact  eclipsing  binary
  (V97).  This binary has a typical W Ursa Majoris light curve.}
\label{eb_plot}
\vspace{0.5cm}
\end{figure} 

\subsection{Classification}

\begin{deluxetable}{lrc}
\tablecaption{Classification Summary}
\tablewidth{0pc}
\tablehead{
\colhead{Variable Type} & \colhead{Count} & \colhead{Period (days)} 
}
\startdata
SX Phoenix         & 1   & 0.05 \\
RR0                & 55  & 0.2-0.4 \\
RR1                & 57  & 0.4-0.8 \\
Eclipsing Binary   & 1   &0.3774\\
Longer Period      & 19   & $P>2$ days\\
\enddata
\end{deluxetable}
\vspace{0.3cm}

We  classified each  variable by  considering  both the  shape of  the
phased light  curve, its amplitude, and its  period.  The relationship
between period and variable type is  seen in Figure 3.  All but one of
the variables  shown in  this histogram are  RR Lyrae  variables.  The
single variable  found with a period  near 0.05 days is  an SX Phoenix
star.  It is  obvious from Figure 3 that there  are two distinct types
of RR  Lyrae stars.  Most  of our detected  variables proved to  be RR
Lyraes which  are then  subdivided into the  two different  types.  We
used the same notation as \citet{cle2001} and classified the RR Lyraes
as  either type  RR0  or  RR1 for  consistency.   Besides the  obvious
difference in periods  and appearance of the phased  light curves (see
Figure  6),  the  distinction  between  RR0 and  RR1  becomes  readily
apparent when comparing the  amplitude of the magnitude fluctuation to
the period of variability as seen in Figure 4.  The shorter-period RR1
variables (lower left) generally have less variability compared to the
longer-period RR0  variables (right).  RR Lyraes  with periods ranging
from 0.2-0.4 are of the type  RR1, and those with periods from 0.4-0.8
are of the type RR0. A summary of the numbers of each type of variable
classified  is shown  in Table  1.  The  majority of  our  newly found
variables  were  of the  RR1  type.  Given  that  RR1's  have a  lower
amplitude of pulsation, it is  reasonable to assume that many of these
would not have been found in previous photographic studies.

The shortest period of our  detected variables (0.049 days) is that of
an SX Phoenix star (V161).   Because of the short term variability and
a longer  term trend  from night  to night, we  found it  difficult to
combined the data over several night  so in figure 6 we only show the
phased  light curve for  a single  night(with about 6 cycles per night)
to eliminate longer term variations.

\setlongtables
\begin{longtable}{rcccccrcc}
\tabletypesize{\scriptsize}
\tablecaption{Variables Found in M14}
\tablewidth{0pt}
\tablehead{
\colhead{V$\#$} & \multicolumn{3}{c}{RA (h,m,s)} & 
\multicolumn{3}{c}{Dec
 ($^{\circ}$,$\arcmin$,$\arcsec$)} &
\colhead{Period (d)} & \colhead{Type}
}
\startdata
 1  & 17 & 37 & 37.22 & -3 & 13 & 59.4 &$\sim$20& lp\\
 2  & 17 & 37 & 28.39 & -3 & 16 & 44.7 & 2.7939 & W Vir\\
 3  & 17 & 37 & 35.87 & -3 & 16 & 14.8 & 0.5223 & RR0\\
 4  & 17 & 37 & 46.87 & -3 & 13 & 32.9 & 0.6514 & RR0\\
 5  & 17 & 37 & 27.02 & -3 & 13 & 16.5 & 0.5488 & RR0\\
 6  & 17 & 37 & 38.37 & -3 & 16 &  3.0 & $>$25 & lp\\
 7  & 17 & 37 & 40.25 & -3 & 16 & 22.0 & 13.361 & lp\\
 8  & 17 & 37 & 42.47 & -3 & 14 & 10.2 & 0.6860 & RR0\\
 9  & 17 & 37 & 46.21 & -3 & 15 & 25.1 & 0.5387 & RR0\\
10  & 17 & 37 & 32.77 & -3 & 18 &  9.9 & 0.5860 & RR0\\
11  & 17 & 37 & 49.14 & -3 & 18 & 28.5 & 0.6044 & RR0\\
12  & 17 & 37 & 51.11 & -3 & 17 & 43.1 & 0.5045 & RR0\\
13  & 17 & 37 & 34.21 & -3 & 16 & 44.2 & 0.5340 & RR0\\
14  & 17 & 37 & 39.63 & -3 & 14 & 48.5 & 0.4722 & RR0\\
15  & 17 & 37 & 27.14 & -3 & 12 & 17.7 & 0.5578 & RR0\\
16  & 17 & 37 & 30.83 & -3 & 15 & 21.3 & 0.6005 & RR0\\
17  & 17 & 37 & 20.90 & -3 & 12 & 42.8 &$\sim$11 & lp\\
18  & 17 & 37 & 40.28 & -3 & 15 &  7.7 & 0.4790 & RR0\\
19  & 17 & 37 & 27.59 & -3 & 14 & 43.3 & 0.5460 & RR0\\
20  & 17 & 37 & 26.40 & -3 & 13 &  7.1 & 0.2635 & RR1\\
21  & 17 & 37 & 40.87 & -3 & 12 & 40.3 & 0.3188 & RR1\\
22  & 17 & 37 & 40.78 & -3 & 13 & 11.0 & 0.6564 & RR0\\
23  & 17 & 37 & 41.31 & -3 & 16 &  4.2 &$>$25 & lp\\
24  & 17 & 37 & 35.97 & -3 & 13 & 30.4 & 0.5199 & RR0\\
29  & 17 & 37 & 31.61 & -3 & 17 & 15.9 &$>$25 & lp\\
30  & 17 & 37 & 41.13 & -3 & 14 & 57.7 & 0.5345 & RR0\\
31  & 17 & 37 & 33.46 & -3 & 14 & 13.7 & 0.6196 & RR0\\
32  & 17 & 37 & 38.42 & -3 & 12 & 18.7 & 0.6559 & RR0\\
33  & 17 & 37 & 26.83 & -3 & 14 & 31.1 & 0.4805 & RR0\\
34  & 17 & 37 & 31.37 & -3 & 14 & 18.6 & 0.6066 & RR0\\
35  & 17 & 37 & 28.42 & -3 & 15 & 34.8 & 0.5266 & RR0\\
36  & 17 & 37 & 50.03 & -3 & 20 & 28.4 & 0.6787 & RR0\\
37  & 17 & 37 & 36.50 & -3 & 14 & 27.1 & 0.4897 & RR0\\
38  & 17 & 37 & 36.81 & -3 & 15 &  2.7 & 0.5084 & RR0\\
39  & 17 & 37 & 39.15 & -3 & 14 & 46.1 & 0.5760 & RR0\\
41  & 17 & 37 & 34.91 & -3 & 14 & 47.1 & 0.2594 & RR1\\
42  & 17 & 37 & 38.52 & -3 & 14 & 33.5 & 0.6313 & RR0\\
43  & 17 & 37 & 40.52 & -3 & 14 & 23.4 & 0.5222 & RR0\\
44  & 17 & 37 & 37.35 & -3 & 12 & 47.6 & 0.2894 & RR1\\
46  & 17 & 37 & 42.12 & -3 & 15 & 50.7 & 0.3330 & RR1\\
47  & 17 & 37 & 30.11 & -3 & 14 & 18.1 & 0.8774 & RR0\\
48  & 17 & 37 & 35.67 & -3 & 14 &  4.7 & 0.4677 & RR0\\
49  & 17 & 37 & 29.66 & -3 & 15 &  3.8 & 0.6423 & RR0\\
51  & 17 & 37 & 43.12 & -3 & 19 & 52.2 & 0.2682 & RR1\\
55  & 17 & 37 & 38.28 & -3 & 12 & 58.8 & 0.3374 & RR1\\
56  & 17 & 37 & 31.63 & -3 & 17 & 48.8 & 0.3413 & RR1\\
57  & 17 & 37 & 45.06 & -3 & 16 & 40.9 & 0.5672 & RR0\\
58  & 17 & 37 & 27.91 & -3 & 15 & 18.5 & 0.4179 & RR1\\
59  & 17 & 37 & 33.97 & -3 & 14 & 15.9 & 0.5570 & RR0\\
60  & 17 & 37 & 38.87 & -3 & 13 & 50.6 & 0.5789 & RR0\\
61  & 17 & 37 & 37.00 & -3 & 15 & 28.6 & 0.5698 & RR0\\
62  & 17 & 37 & 20.68 & -3 & 17 & 19.4 & 0.6380 & RR0\\
70  & 17 & 37 & 38.93 & -3 & 15 &  8.1 & 0.6060 & RR0\\
73  & 17 & 37 & 36.39 & -3 & 14 & 38.1 & $>$25 & lp\\
74  & 17 & 37 & 36.54 & -3 & 13 & 14.5 & $>$25 & lp\\
75  & 17 & 37 & 38.43 & -3 & 14 & 55.7 & 0.5453 & RR0\\
76  & 17 & 37 & 29.05 & -3 & 14 & 44.1 & 1.8979 & RR0\\
77  & 17 & 37 & 28.81 & -3 & 13 & 46.5 & 0.7910 & RR0\\
78  & 17 & 37 & 26.97 & -3 & 14 & 47.4 & 0.3102 & RR1\\
79  & 17 & 37 & 35.35 & -3 & 15 & 00.3 & 0.5597 & RR0\\
88  & 17 & 37 & 30.85 & -3 & 14 & 32.4 & 0.3130 & RR1\\
90  & 17 & 37 & 33.51 & -3 & 15 & 16.2 & 0.3512 & RR1\\
94  & 17 & 37 & 23.04 & -3 & 14 & 50.5 & 0.2585 & RR1\\
95  & 17 & 37 & 24.17 & -3 & 17 & 28.5 & 0.3596 & RR1\\
96  & 17 & 37 & 24.72 & -3 & 14 & 46.3 & 0.2524 & RR1\\
97  & 17 & 37 & 25.02 & -3 & 18 & 36.5 & 0.3774 & EB\\
98  & 17 & 37 & 25.81 & -3 & 12 & 47.8 & 0.2579 & RR1\\
99  & 17 & 37 & 26.58 & -3 & 14 & 46.1 & $\sim$15 & lp\\
100 & 17 & 37 & 29.72 & -3 & 15 & 20.5 & 0.2612 & RR1\\
101 & 17 & 37 & 30.28 & -3 & 15 & 52.7 & $>$25 & lp\\
102 & 17 & 37 & 31.38 & -3 & 16 &  0.6 & $\sim$15 & lp\\
103 & 17 & 37 & 32.48 & -3 & 14 & 12.5 & $>$25 & lp\\
104 & 17 & 37 & 32.69 & -3 & 14 & 56.0 & 0.2620 & RR1\\
105 & 17 & 37 & 32.95 & -3 & 14 &  3.3 & 0.2793 & RR1\\
106 & 17 & 37 & 33.28 & -3 & 14 & 46.4 & 0.5465 & RR0\\
107 & 17 & 37 & 33.56 & -3 & 14 & 48.2 & 0.2950 & RR1\\
108 & 17 & 37 & 33.62 & -3 & 15 & 12.0 & 0.3565 & RR1\\
109 & 17 & 37 & 33.64 & -3 & 14 & 40.9 & 0.6562 & RR0\\
110 & 17 & 37 & 33.64 & -3 & 16 & 10.2 & 0.3013 & RR1\\
111 & 17 & 37 & 33.71 & -3 & 15 & 11.9 & 0.2791 & RR1\\
112 & 17 & 37 & 33.76 & -3 & 17 & 14.7 & 0.3156 & RR1\\
113 & 17 & 37 & 33.91 & -3 & 14 & 27.5 & 0.2574 & RR1\\
114 & 17 & 37 & 33.91 & -3 & 14 & 52.5 & 0.3325 & RR1\\
115 & 17 & 37 & 33.95 & -3 & 14 & 24.1 & 0.3359 & RR1\\
116 & 17 & 37 & 34.03 & -3 & 15 & 37.7 & 0.2521 & RR1\\
117 & 17 & 37 & 34.10 & -3 & 14 & 35.9 & 0.3394 & RR1\\
118 & 17 & 37 & 34.24 & -3 & 16 & 13.0 & 0.3803 & RR1\\
119 & 17 & 37 & 34.45 & -3 & 14 & 52.1 & 0.3341 & RR1\\
120 & 17 & 37 & 34.65 & -3 & 13 & 29.6 & 0.3279 & RR1\\
121 & 17 & 37 & 35.04 & -3 & 15 & 19.7 & 0.2704 & RR1\\
122 & 17 & 37 & 35.30 & -3 & 14 & 39.2 & 0.5590 & RR0\\
123 & 17 & 37 & 35.55 & -3 & 15 & 44.1 & 0.2846 & RR1\\
124 & 17 & 37 & 35.75 & -3 & 15 & 27.3 & 0.2762 & RR0\\
125 & 17 & 37 & 35.79 & -3 & 15 & 15.1 & $\sim$20 & lp\\
126 & 17 & 37 & 35.83 & -3 & 14 & 53.7 & 0.2939 & RR1\\
127 & 17 & 37 & 35.92 & -3 & 14 & 32.8 & 0.2983 & RR1\\
128 & 17 & 37 & 35.95 & -3 & 13 & 52.8 & 0.3935 & RR1\\
129 & 17 & 37 & 36.11 & -3 & 15 &  1.4 & 0.2796 & RR1\\
130 & 17 & 37 & 36.26 & -3 & 15 & 25.4 & 0.5961 & RR0\\
131 & 17 & 37 & 36.27 & -3 & 14 & 54.3 & 0.2697 & RR1\\
132 & 17 & 37 & 36.53 & -3 & 15 & 14.7 & 0.4822 & RR0\\
133 & 17 & 37 & 36.64 & -3 & 18 & 15.9 & 0.3030 & RR1\\
134 & 17 & 37 & 36.69 & -3 & 15 &  5.8 & 0.5072 & RR0\\
135 & 17 & 37 & 36.82 & -3 & 13 & 41.9 & 0.3382 & RR1\\
136 & 17 & 37 & 36.83 & -3 & 15 & 24.5 & 0.3305 & RR1\\
137 & 17 & 37 & 37.48 & -3 & 14 & 40.7 & 0.2763 & RR1\\
138 & 17 & 37 & 37.93 & -3 & 15 & 33.0 & 0.3710 & RR1\\
139 & 17 & 37 & 37.99 & -3 & 17 & 23.7 & 0.2722 & RR1\\
140 & 17 & 37 & 38.21 & -3 & 15 & 51.9 & 0.7609 & RR0\\
141 & 17 & 37 & 38.31 & -3 & 14 & 27.2 & 0.6433 & RR0\\
142 & 17 & 37 & 38.72 & -3 & 14 &  2.1 & 0.4618 & RR0\\
143 & 17 & 37 & 38.82 & -3 & 16 & 31.2 & 0.3185 & RR1\\
144 & 17 & 37 & 39.27 & -3 & 14 & 29.7 & 0.3758 & RR1\\
145 & 17 & 37 & 39.97 & -3 & 15 &  1.4 & 0.3003 & RR1\\
146 & 17 & 37 & 40.16 & -3 & 16 &  8.6 & $>$25 & lp\\
147 & 17 & 37 & 40.22 & -3 & 15 & 55.5 & 0.4932 & RR0\\
148 & 17 & 37 & 40.73 & -3 & 17 &  0.3 & 0.2651 & RR1\\
149 & 17 & 37 & 41.07 & -3 & 10 &  4.7 & 0.5523 & RR0\\
150 & 17 & 37 & 41.14 & -3 & 14 &  8.0 & 0.8028 & RR0\\
151 & 17 & 37 & 41.19 & -3 & 14 & 22.7 & 0.3198 & RR1\\
152 & 17 & 37 & 41.90 & -3 & 15 & 37.1 & $>$25 & lp\\
153 & 17 & 37 & 41.98 & -3 & 11 & 56.0 & 0.2650 & RR1\\
154 & 17 & 37 & 42.54 & -3 & 13 & 59.4 & 0.2551 & RR1\\
155 & 17 & 37 & 42.93 & -3 & 14 &  5.3 & $\sim$15 & lp\\
156 & 17 & 37 & 43.60 & -3 & 14 & 24.6 & $>$25 & lp\\
157 & 17 & 37 & 43.79 & -3 & 16 & 13.4 & 0.2626 & RR1\\
158 & 17 & 37 & 43.88 & -3 & 13 &  0.7 & 0.7242 & RR0\\
159 & 17 & 37 & 43.99 & -3 & 13 & 44.8 & 0.2897 & RR1\\
160 & 17 & 37 & 44.24 & -3 & 15 & 34.8 & 0.3400 & RR1\\
161 & 17 & 37 & 44.52 & -3 & 11 & 51.5 & 0.0490 & SX Phe\\
162 & 17 & 37 & 49.69 & -3 & 11 & 49.2 & $\sim$20 & lp\\
163 & 17 & 37 & 53.71 & -3 & 14 & 18.8 & 0.2903 & RR1\\
164 & 17 & 37 & 56.28 & -3 & 10 & 16.3 & 0.3326 & RR1
\enddata
\end{longtable}
\tablenotemark{We assign 'lp' to any variable with a period greater than
  3 days.}

\begin{figure*}
\epsscale{1.15}
\plotone{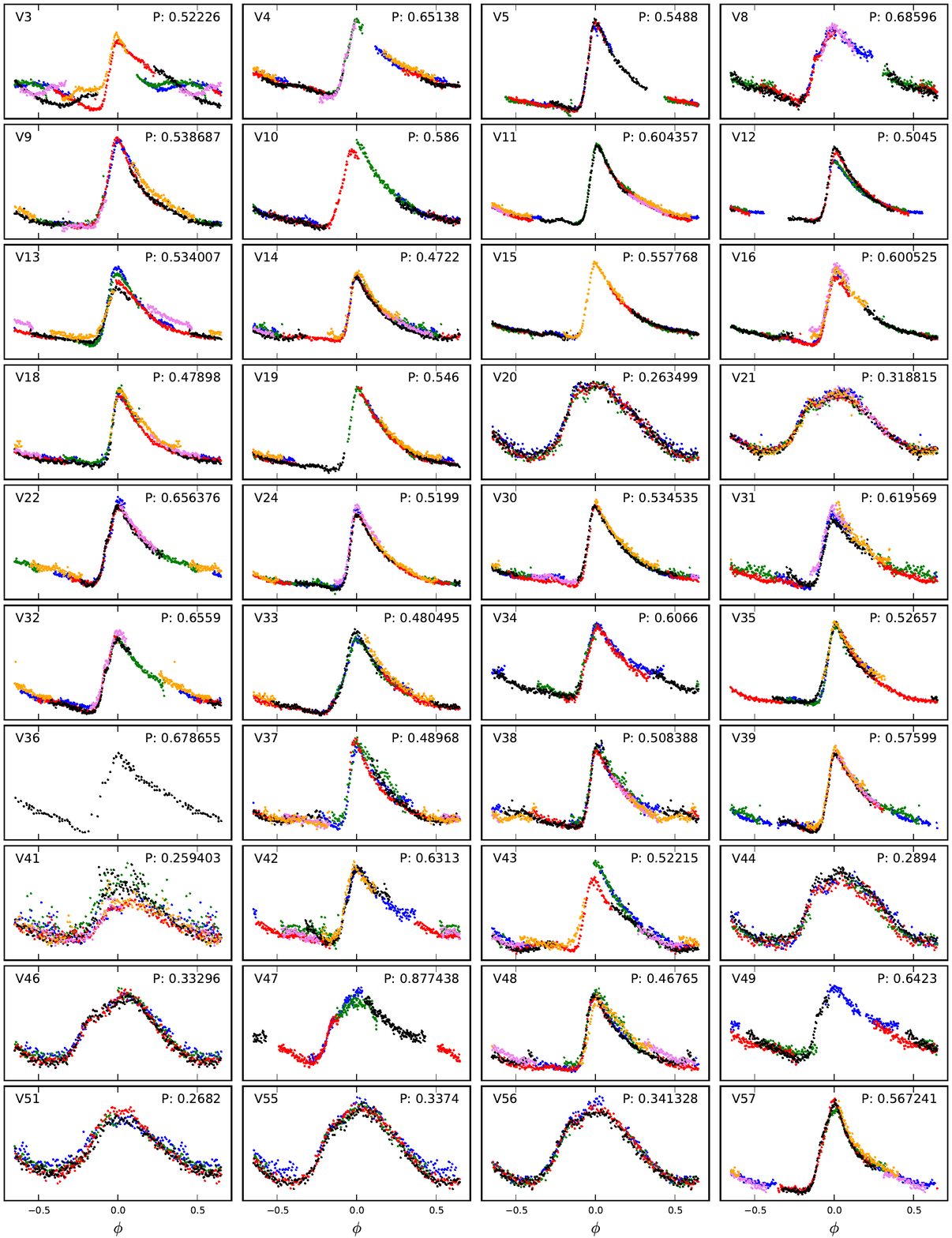}
\caption{Phased light  curves for all detected  variables with periods
  under 3  days.  The  vast majority of  these variables are  RR Lyrae
  stars.}
\label{phased_variables}
\end{figure*}

\begin{figure*}
\epsscale{1.15}
\setcounter{figure}{5}
\plotone{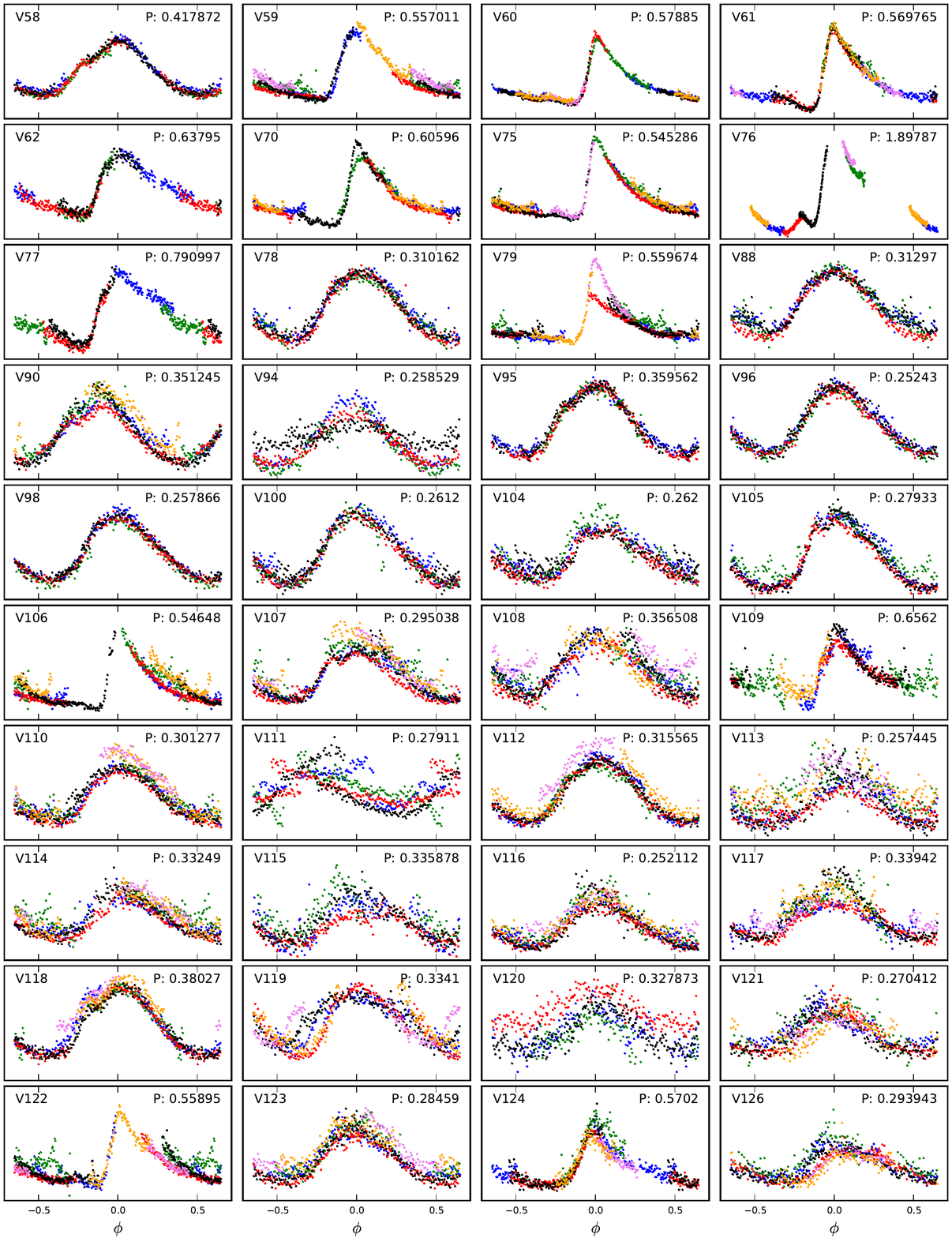}
\caption{(Continued)  Phased light curves  for  all detected  variables
  with periods under 3 days.  The vast majority of these variables are
  RR Lyrae stars.}\label{phased_variables_2}
\end{figure*}

\begin{figure*}
\epsscale{1.15}
\plotone{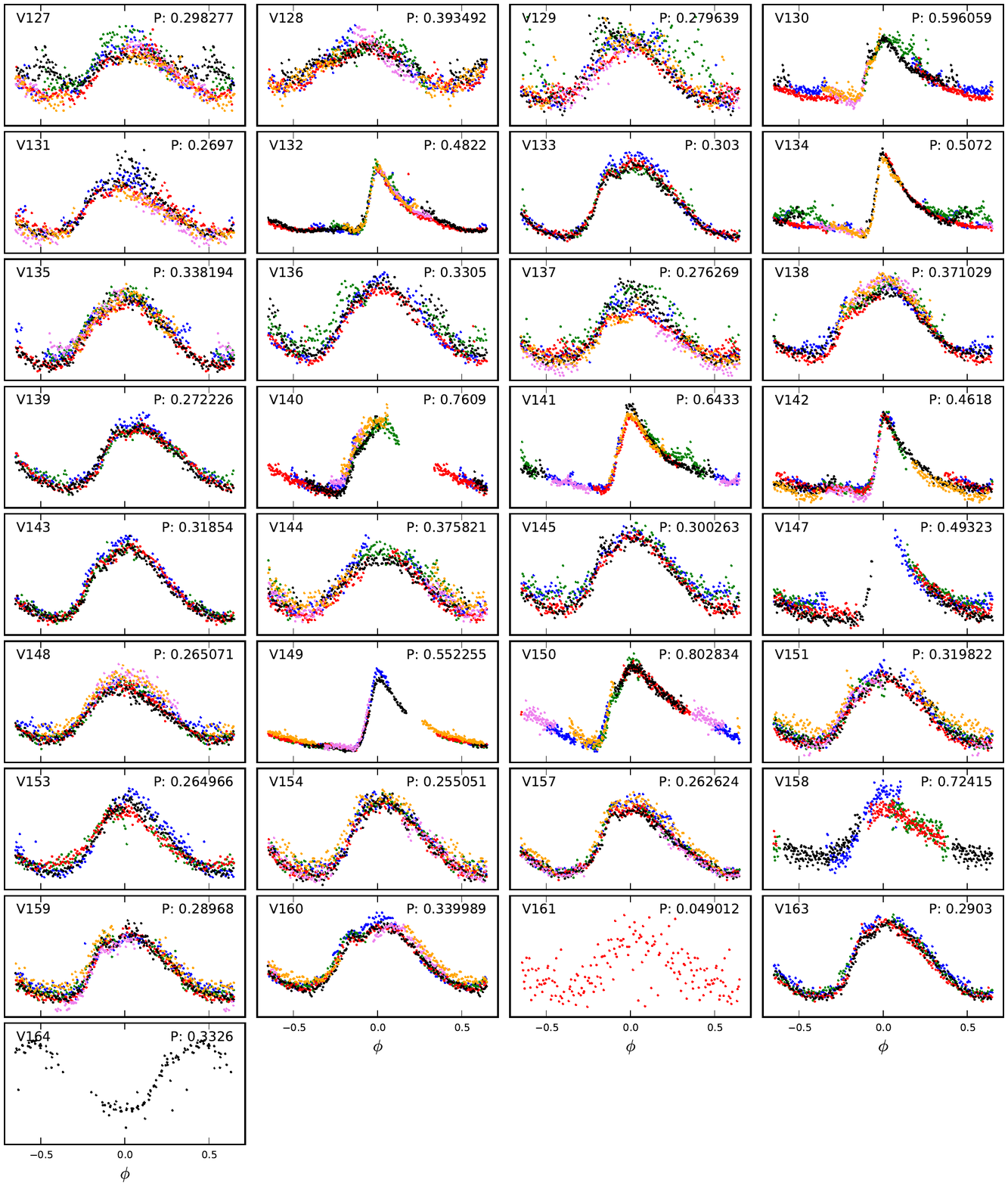}
\setcounter{figure}{5}
\caption{(Continued)  Phased light curves  for  all detected  variables
  with periods under 3 days.  The vast majority of these variables are
  RR Lyrae stars.}\label{phased_variables_3}
\end{figure*}

\begin{figure*}[ht]
\epsscale{1.15}
\plotone{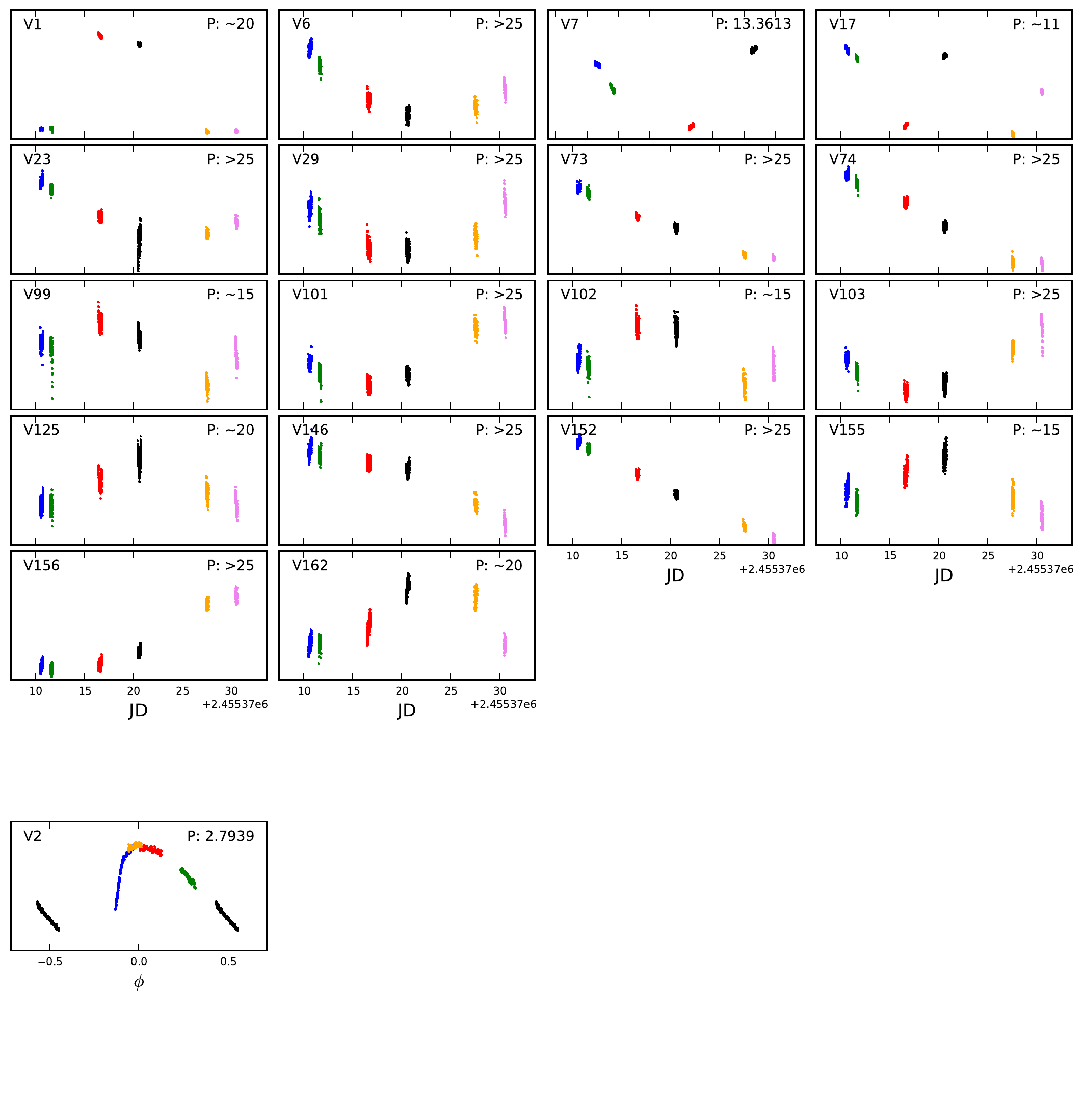}
\caption{Light curves for all detected variables with periods over 2
  days.}
\label{longterm_variables}
\end{figure*}

Several of  the RR0  variables showed evidence  of the  Blazhko effect
\citep{blazhko07,smith04}  resulting in  long term  modulation  in the
amplitude.   Although our observations  were not  long enough  to find
this secondary  period, the effect is  apparent in V3,  V13, V31, V43,
V70, and V79 (Figure \ref{phased_variables}).

Among the new variables we  detected was an  eclipsing binary shown in
Figure \ref{eb_plot}, V97.   The shape of the light  curve indicates it
is a W UMa contact  binary.  The distortion between and similar depths
of  the eclipses  suggest an  over-contact binary,  while  the relative
flatness    of     eclipses    suggest    a     large    mass    ratio
\citep{rucinski92,webbink03}.

By running all of our Cerro Tololo data through ISIS simultaneously we
were able  to detect  several possible long  term variables.   Many of
these  are  most  likely   red  giant  stars  undergoing  longer  term
pulsations.  All of our potential longer period variables are included
in Table 2 listed as 'lp'  and their light curves are plotted unphased
in Figure  \ref{longterm_variables}.  Our observing time  span was not
long  enough to observe  multiple cycles  for these  variables.  Those
with periods listed are only estimates.

All detected variables and  their periods and classification are shown
in Table 2.   Light curves of variables with shorter  periods are shown
phased in Figure 6, whereas the light curves of longer period variables
are shown unphased except for V2 in Figure 7. V2 is an W Vir star with
a period of  2.7939 days.  Given its relatively  short period, we show
its light curve as phased.
\vspace*{.4cm}

\section{Conclusions}
We observed the globular star cluster  M14 for 12 nights over a 40 day
period using  the SARA telescopes located  at KPNO and  CTIO.  We used
the image subtraction method of \citet{ala2000} to search for variable
stars in  M14.  We confirmed 63 previously  known variables catalogued
by Wehlau  \& Froelich  (1994).  In addition  to the  previously known
variables we  have identified 71  new variables.  Of the  variables we
were  able to  detect we  have confirmed  the periods  the  Wehlau and
Froelich RR  Lyrae stars with one  exception.  Of the  total number of
confirmed variables we found 55 RR0, 57 RR1, 19 variables with periods
greater than 2  days, a W UMa contact binary, and  an SX Phoenix star.
We confirmed the periods of  previously found variables as well as the
determined  of the periods,  classification, and  light curves  of the
newly discovered variables.




\vspace*{-.4cm}
\acknowledgments

We thank  C.  Alard  for making ISIS  2.2 publically  available.  This
project  was  funded  by  the  National  Science  Foundation  Research
Experiences  for  Undergraduates   (REU)  program  through  grant  NSF
AST-1004872. Additionally Z. Liu  and B.  Murphy were partially funded
by the Butler Insitute for Research and Scholarship.  The authors also
thank F.   Levinson for a generous gift  enabling Butler University's
membership in the SARA consortium.






\begin{thebibliography}{}

\bibitem[Alard  \&   Lupton(1998)]{ala1998}  Alard,  C.,   \&  Lupton,
  R.~H.\ 1998, \apj, 503, 325
\bibitem[Alard(2000)]{ala2000} Alard, C.\ 2000, \aaps, 144, 363
\bibitem[Bla{\v   z}ko(1907)]{blazhko07}   Bla{\v   z}ko,  S.\   1907,
  Astronomische Nachrichten, 175, 325
\bibitem[Clement et al.(2001)]{cle2001}  Clement, C.~M., et al.\ 2001,
  \aj, 122, 2587
\bibitem[Harris(1996)]{1996AJ....112.1487H} Harris,  W.~E.\ 1996, \aj,
  112, 1487
\bibitem[Sawyer-Hogg  \& Wehlau(1968)]{sawyerhogg68} Sawyer  Hogg, H.,
  \&   Wehlau,  A.~W.\   1968,  Publications   of  the   David  Dunlap
  Observatory, 2, 493
\bibitem[Jacobs(2004)]{2004,  unpublished thesis} Jacobs,  C.L.\ 2004,
  unpublished undergraduate thesis, Duke University
\bibitem[Rucinski(1992)]{rucinski92} Rucinski,  S.~M.\ 1992, The Realm
  of Interacting Binary Stars, 177, 111
\bibitem[Scargle(1982)]{scargle82}  Scargle, J.~D.\  1982,  \apj, 263,
  835
\bibitem[Smith(2004)]{smith04} Smith, H.~A.\  2004, RR Lyrae Stars, by
  Horace   A.~Smith,   ~ISBN   0521548179.~Cambridge,  UK:   Cambridge
  University Press
\bibitem[Webbink(2003)]{webbink03}  Webbink, R.~F.\  2003,  3D Stellar
  Evolution, 293, 76
\bibitem[Wehlau \&  Froelich(1994)]{weh1994} Wehlau, A.,  \& Froelich,
  N.\ 1994, \aj, 108, 134

\end{thebibliography}
\end{document}